\documentclass[sigconf]{acmart}
\usepackage{amsthm}
\usepackage{subcaption}
\usepackage{paralist}
\usepackage{bm}
\usepackage{amsfonts}
\usepackage{multirow}
\usepackage{cases}
\usepackage{booktabs}
\usepackage{threeparttable}
\usepackage{epstopdf}
\usepackage{upgreek}
\usepackage{endnotes}
\usepackage{etoolbox}
\usepackage{algpseudocode}
\usepackage{amsmath}
\usepackage{algorithm}
\usepackage{xspace}
\usepackage{array}
\usepackage{enumitem}
\usepackage{hyperref}
\usepackage{balance}

\AtBeginDocument{%
  \providecommand\BibTeX{{%
    \normalfont B\kern-0.5em{\scshape i\kern-0.25em b}\kern-0.8em\TeX}}}
\newcommand{\eat}[1]{}
\newcommand{\paratitle}[1]{\vspace{1.5ex}\noindent \textbf{#1}}
\newcommand{\ie}{\emph{i.e.,}\xspace}
\newcommand{\eg}{\emph{e.g.,}\xspace}

\newcommand{\baby}{\textsc{Conde}\xspace}

\copyrightyear{2021} 
\acmYear{2021} 
\setcopyright{acmcopyright}\acmConference[CIKM '21]{Proceedings of the 30th ACM International Conference on Information and Knowledge Management}{November 1--5, 2021}{Virtual Event, QLD, Australia}
\acmBooktitle{Proceedings of the 30th ACM International Conference on Information and Knowledge Management (CIKM '21), November 1--5, 2021, Virtual Event, QLD, Australia}
\acmPrice{15.00}
\acmDOI{10.1145/3459637.3482417}
\acmISBN{978-1-4503-8446-9/21/11}

\begin{document}
\fancyhead{}
\title{Concept-Aware Denoising Graph Neural Network \\ for Micro-Video Recommendation}\thanks{$^\dagger$Chenliang Li is the corresponding author. Work done when Yiyu, Qian and Yu were interns at Kuaishou.}
\author{Yiyu Liu$^{1}$, Qian Liu$^{1}$, Yu Tian$^{1}$, Changping Wang$^{2}$, Yanan Niu$^{2}$, Yang Song$^{2}$, Chenliang Li$^{1\dagger}$}
\affiliation{
\institution{
  $^1$Key Laboratory of Aerospace Information Security and Trusted Computing, Ministry of Education, School of Cyber Science and Engineering, Wuhan University, Wuhan, 430072, China \\
  \{liuyiyu,littleliu\}@whu.edu.cn; s.braylon1002@gmail.com; cllee@whu.edu.cn\\
  $^2$Kuaishou Technology Co., Ltd., Beijing, 10010, China\\\{wangchangping,niuyanan,yangsong\}@kuaishou.com
  }
  \country{}
}

\def\authors{Yiyu Liu, Qian Liu, Yu Tian, Changping Wang, Yanan Niu, Yang Song, Chenliang Li}
\renewcommand{\shortauthors}{Yiyu Liu, et al.}

\begin{abstract}
Recently, micro-video sharing platforms such as Kuaishou and Tiktok have become a major source of information for people's lives. Thanks to the large traffic volume, short video lifespan and streaming fashion of these services, it has become more and more pressing to improve the existing recommender systems to accommodate these challenges in a cost-effective way. In this paper, we propose a novel concept-aware denoising graph neural network (named \baby) for micro-video recommendation. \baby consists of a three-phase graph convolution process to derive user and micro-video representations: \textit{warm-up propagation, graph denoising} and \textit{preference refinement}. A heterogeneous tripartite graph is constructed by connecting user nodes with video nodes, and video nodes with associated concept nodes, extracted from captions and comments of the videos. To address the noisy information in the graph,  we introduce a user-oriented graph denoising phase to extract a subgraph which can better reflect the user's preference. Despite the main focus of micro-video recommendation in this paper, we also show that our method can be generalized to other types of tasks. Therefore, we also conduct empirical studies on a well-known public E-commerce dataset. The experimental results suggest that the proposed \baby achieves significantly better recommendation performance than the existing state-of-the-art solutions.  
\end{abstract}



\maketitle
\section{Introduction}
Information overload has become an increasingly crucial challenge for today's world. 
In the past few years, micro-video sharing platforms like Kuaishou have harnessed a huge user base on a global scale. Users on these platforms can share micro-videos ranging from a few seconds to minutes with their friends and the public. Due to the significant cognitive load reduction empowered by visual communication, micro-videos have become the new time killer. As of early $2020$, there are more than $300$M daily active users (DAUs) and over $20$B videos on the Kuaishou platform.
 
 Faced with a massive upsurge of micro-video data traffic, the key techniques to alleviate the information overload problem is recommender systems, which aim to precisely rank items in terms of the user's preference. A myriad of recommendation algorithms have been proposed in the past, including sequential based recommendation~\cite{MIMN/KDD2019,aliSIM/2020}, graph-based recommendation ~\cite{pinSage/KDD18, pinnersage/KDD20} and a range of CTR prediction models that are practical in industrial settings~\cite{FM,FFM,NFM,AFM,PNN,FNN}.

 Nevertheless, the following three unique characteristics in micro-video sharing platforms impede the existing recommendation techniques to deliver good performance: 

\textbf{C1:} The existing multi-modal models that extract visual content are infeasible for micro-video sharing platforms due to the continuous and large traffic volume nature of the latter. On the other hand, exploiting the rich story expressed in a micro-video will help understand the user's preference to its maximum.
 
\textbf{C2:} Although micro-video platforms provide users with "like" and "comment" buttons to interact with the video, the vast majority of user actions are still just swiping up to the next video, leaving very sparse user feedback. Without an explicit user interaction, it is hard to tell whether the user really likes what she/he watches. Although we can utilize watching time to speculate the user's preference (\ie analogous to click behavior), there would still be many false positives which hinder the effective preference learning for recommendation.
 
\textbf{C3:} The average lifecycle of a micro-video is extremely short. In our data analysis over user behaviors from a large scale micro-video sharing platform, it is observed that the number of user interactions over a micro-video reduces sharply just two days later after its announcement\footnote{Due to the privacy policy of the company, we cannot release the lifecycle distribution here.}. Moreover, most of the user behaviors fall on the micro-videos uploaded by a few Internet celebrities. Specifically, the micro-videos that have less than $10$ user clicks comprise about $85\%$ in our dataset. This number goes up to $96\%$ when we use $50$ as the cutoff threshold.
These long-tail ones are rarely recommended. The feedback loop underlying the recommender system further aggravates this recommendation bias, which is called "Matthew Effect"~\cite{esam/sigir20}.

\begin{figure}[!]
    \centering
  \begin{subfigure}[b]{0.23\textwidth}
    \includegraphics[width=0.9\textwidth]{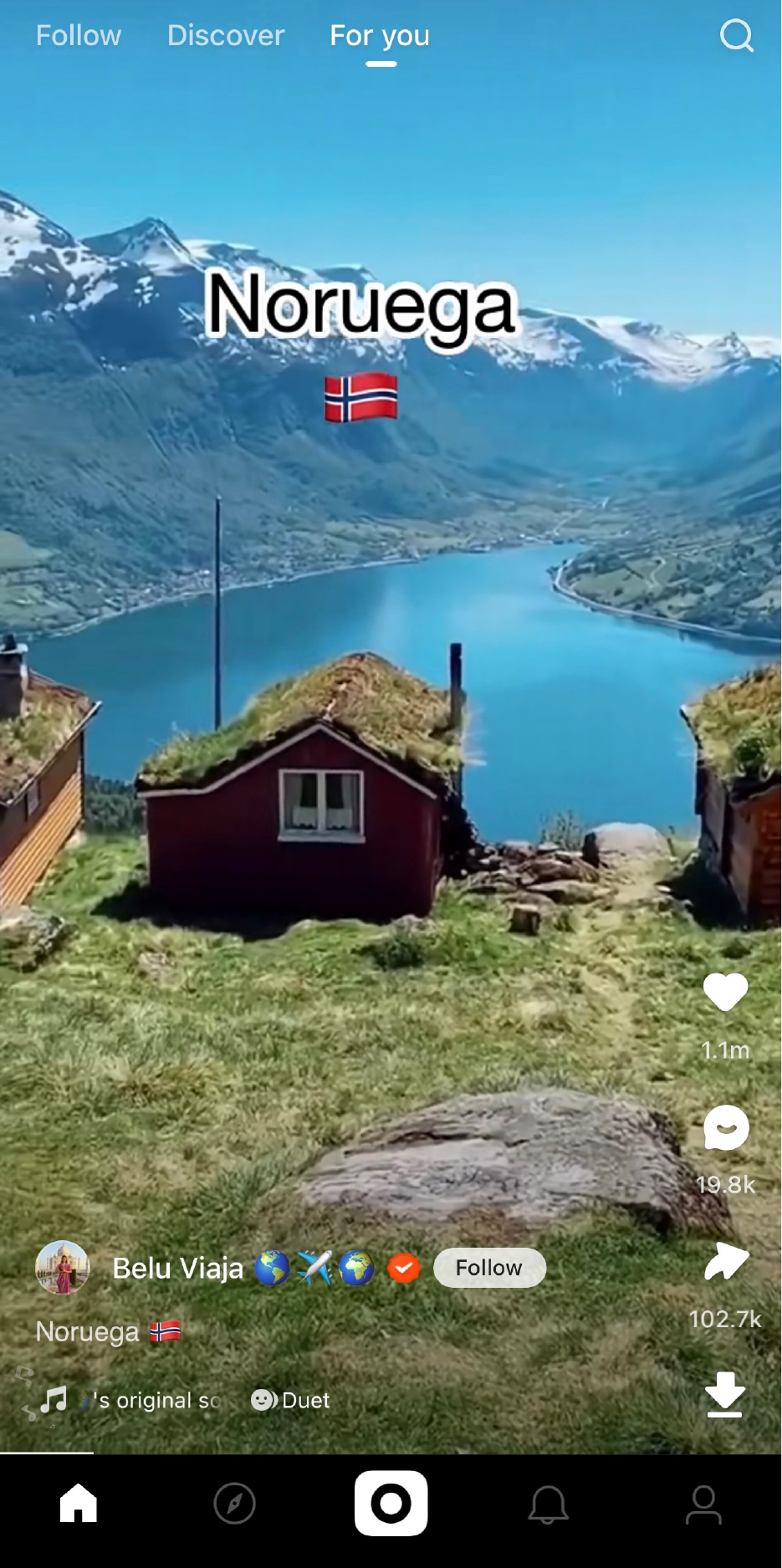}
    \caption{The homepage of Kwai.}
    \label{si2D1}
  \end{subfigure}
   \begin{subfigure}[b]{0.23\textwidth}
    \includegraphics[width=0.9\textwidth]{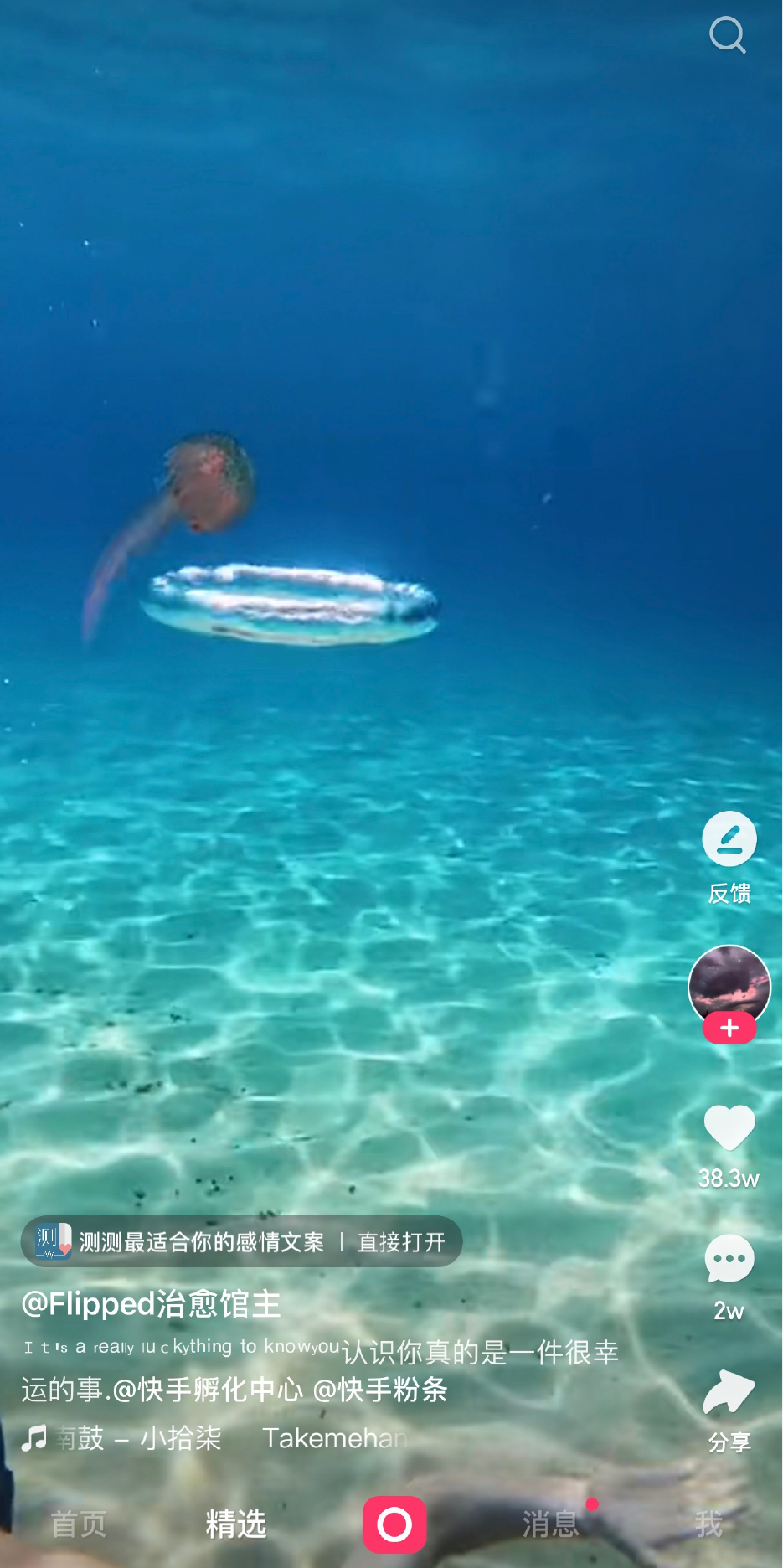}
    \caption{The homepage of Kuaishou.}
    \label{si2D2}
    \end{subfigure}
    \includegraphics[width=0.4\textwidth]{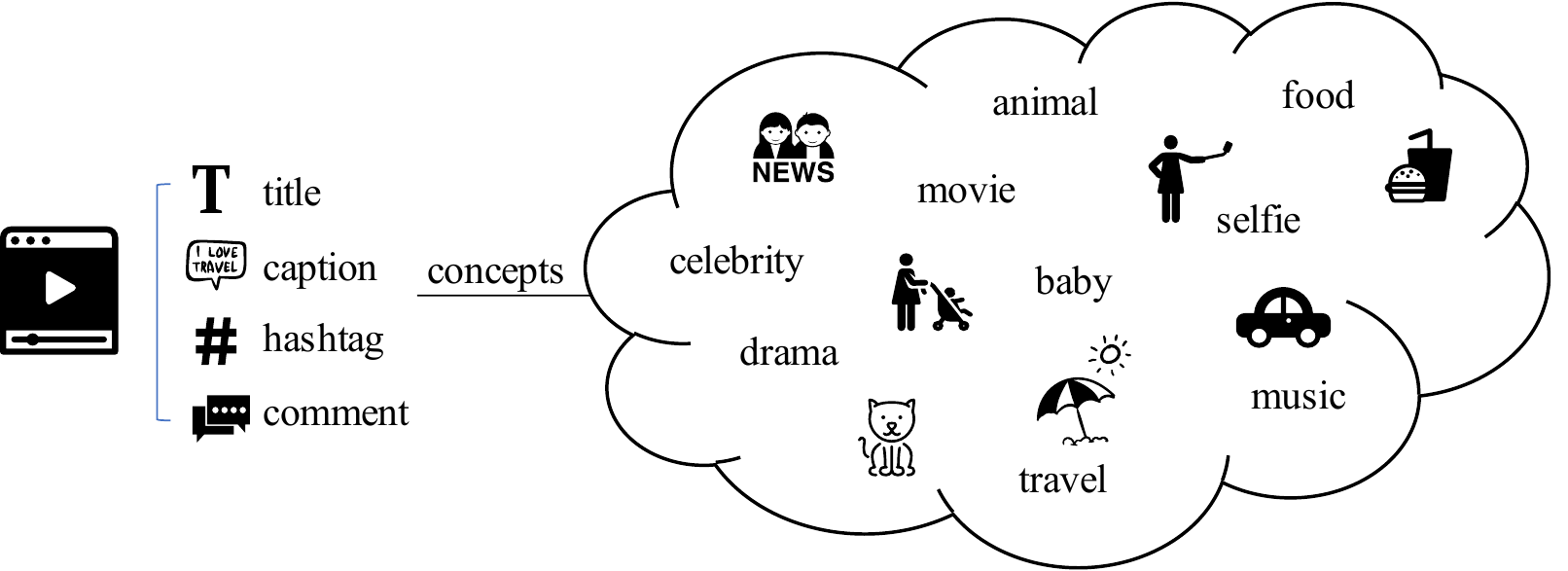}
    \centering
    \caption{A snapshot of the popular micro-video platform kuaishou and its international version Kwai. The text information within a micro-video contains rich semantics, which can be extracted as millions of concepts in our work.}
\label{fig:kwaiPage}
\end{figure}

All these characteristics together give rise to sparser user interactions and poor performance for micro-video recommendation, especially for long-tail ones. To this end, in this paper, we propose a novel \textbf{con}cept-aware \textbf{de}noising graph neural network (named \baby) to address the above challenges. In \baby, we aim to build a marriage between textual information and graph neural network to support video content extraction and user preference learning. As illustrated in Figure~\ref{fig:kwaiPage}, the caption and comments associated with a micro-video serve as a good proxy for the video content. Hence, to account for the rich semantics in a micro-video, we resort to extracting concepts mentioned in the captions and the comments of micro-videos. Here, the concepts are defined as the named entities and semantic key phrases (\eg ``heartwarming'', ``classic HK movie''). By connecting the micro-videos to the associated concepts and users to their consumed micro-videos, we can form a tripartite heterogeneous graph, in which the concepts work as a backbone with rich semantic information.

It is intuitive that not all concepts of a micro-video can reflect all different users' preferences equally. Moreover, as aforementioned, there would be many false positives for user-video edges in the graph. Hence, we further introduce a personalized denoising procedure for each user when we derive her/his representation via graph convolution. Specifically, \baby utilizes a three-phase graph convolution process to derive user and item representations. In the \textit{warm-up propagation} phase, we firstly apply convolution by propagating semantic concept information to the related micro-videos and users through the aggregation order: concept $\to$ micro-video $\to$ user. Afterwards, we further perform another layer of convolution over the user-video edges to exploit the collaborative signals. With the user and item representations calculated by the warm-up phase, \textit{graph denoising} phase aims to extract a subgraph for a given user where the noisy micro-videos, concepts and other users in the neighborhood are removed in a breadth-first search fashion. It is expected that the resultant subgraph could reflect the user's preference more precisely. At last, in \textit{preference refinement} phase, we again perform the same convolution as in the warm-up phase over the subgraph to derive the user representation. 

The extensive experiments over a large scale internal micro-video dataset and a traditional E-Commerce dataset in different languages demonstrate that the proposed \baby achieves significantly superior recommendation performance to the existing state-of-the-art technical alternatives. In summary, the main contributions of this paper are as follows:

\begin{itemize}
    \item We fully exploit textual information to support video content extraction. The rich semantics in micro-videos demonstrate great potential in representing the user's preference.
    \item We propose a denoising phase for graph neural networks to help recommendation systems, especially those in a streaming fashion, get rid of noisy information and catch user's highly dynamic preference more precisely.
    \item To the best of our knowledge, this is the first attempt in recommendation systems to study the heterogeneous graph neural network with a denoising purpose.
\end{itemize}

\section{Related Work}
Since our work is related to micro-video recommendation, graph-based recommendation and denoising for graph and recommendation, we therefore mainly focus on reviewing existing methods in these three lines.
\subsection{Micro-Video Recommendation}
Different from item purchase in E-Commerce sites, micro-videos contain much more storylines. To enable better semantic learning for micro-videos, MMGCN~\cite{mmgcn} incorporates multi-modal information into a collaborative filtering framework. It captures modal-specific user preferences by separately constructing a user-item bipartite graph for each modality. ALPINE~\cite{ALPINE/MM2019} also designs a graph-based sequential network for recommendation tasks, which can better model a user's diverse and dynamic interest. THACIL~\cite{THACIL/MM2018} proposes a hierarchical attention mechanism for modeling both short-term and long-term behaviors. Though encouraging performance is achieved by these efforts, they all require visual feature extraction. However, this heavy treatment could be too expensive to be adopted for a real-world application.

\subsection{Graph-based Recommendation}
Recent years have witnessed the superior performance of graph neural networks in many fields that require network embedding and structure modeling. In practice, GNN methods have more expressive capacity than traditional feature-based methods.
The key idea of GNN is to recursively aggregate information from local neighborhoods. Graph Convolution Networks~\cite{GCN/iclr201} is the first work that introduces the convolution operation over a network structure, which gathers the information from source nodes' one-hop neighbors via the neighborhood aggregation, and achieves message passing by stacking multiple GCN layers. Graph Attention Network (GAT)~\cite{GAT/iclr2018} introduces the attention mechanism into GNNs, which enables the model to learn the importance of different neighbors during the information aggregation. These solutions are mainly devised for homogeneous graphs where all nodes or edges are of the same type. To deal with heterogeneous graph that contains different types of nodes and edges, HAN
~\cite{HAN/www2019} extends the original GAT model with a semantic attention layer to learn the importance of different meta-paths. The major idea underlying these works has been widely adopted for different kinds of recommendation tasks.

Specifically, a heterogeneous graph is generally built by considering users, items and relevant entities (\eg side information) in a knowledge graph as nodes, and users' behaviors like click, purchase, add-to-cart and other semantic relations as edges. Based on the real recommendation scenario, \cite{MAGNN,MEIRec,GATNE,HetGNN} extend the recommendation framework as a heterogeneous graph modeling mainly by aggregating the neighborhood information via meta-paths. GraphSAGE\cite{graphSage/nips17} extends the standard graph convolution network to the inductive setting, and utilize batch training and neighborhood sampling to support recommendation on large-scale graphs. AGNN~\cite{qian2019solving} designs an attribute graph based on user-item interactions and utilizes VAE to learn the distribution of attributes for cold-start users/items.
PinSAGE~\cite{pinSage/KDD18} combines random walks and graph convolution layers to aggregate and propagate neighborhood information. MG-BERT~\cite{MG-BERT} utilizes the transformer architecture on the item homogeneous graph to pretrain the item representation, including graph structure reconstruction and masked node feature reconstruction.
With rich semantic information in a knowledge graph, many efforts are devoted to explore auxiliary connections between users and items, therefore making recommendations more accurate and explainable. RippleNet~\cite{ripplenet} is a path-based method that first assigns entities in the KG with initial embeddings and then samples ripple sets from the KG based on the user's historically clicking items. It uses attention networks to simulate user preferences on sampled ripple sets to represent a user. Some of its extensions~\cite{KGAT/KDD2019,akge,KNI/KDD2019} focus on using the embedding propagation mechanism on the item Knowledge Graph. DKN~\cite{dkn} is another representative work that incorporates knowledge graph representation into news recommendations.

\subsection{Graph Denoising}
Performing data denoising in a task-dependent fashion is a promising strategy to alleviate the adverse impact of noisy information. For example, kicking out task-irrelevant edges have been validated to enhance the performance of node classification~\cite{PTDnet/WSDM2019,robust/ICML2019}. Very recently, \cite{sigir21:qin} chose to remove the irrelevant historical records for better sequential recommendation. \cite{sigir21:tian} perform knowledge pruning iteratively by removing irrelevant triples covered by the knowledge graph for better news recommendation.

\section{Method}\label{sec:algo}
\begin{figure}[h]
\centering
\includegraphics[width=0.38\textwidth]{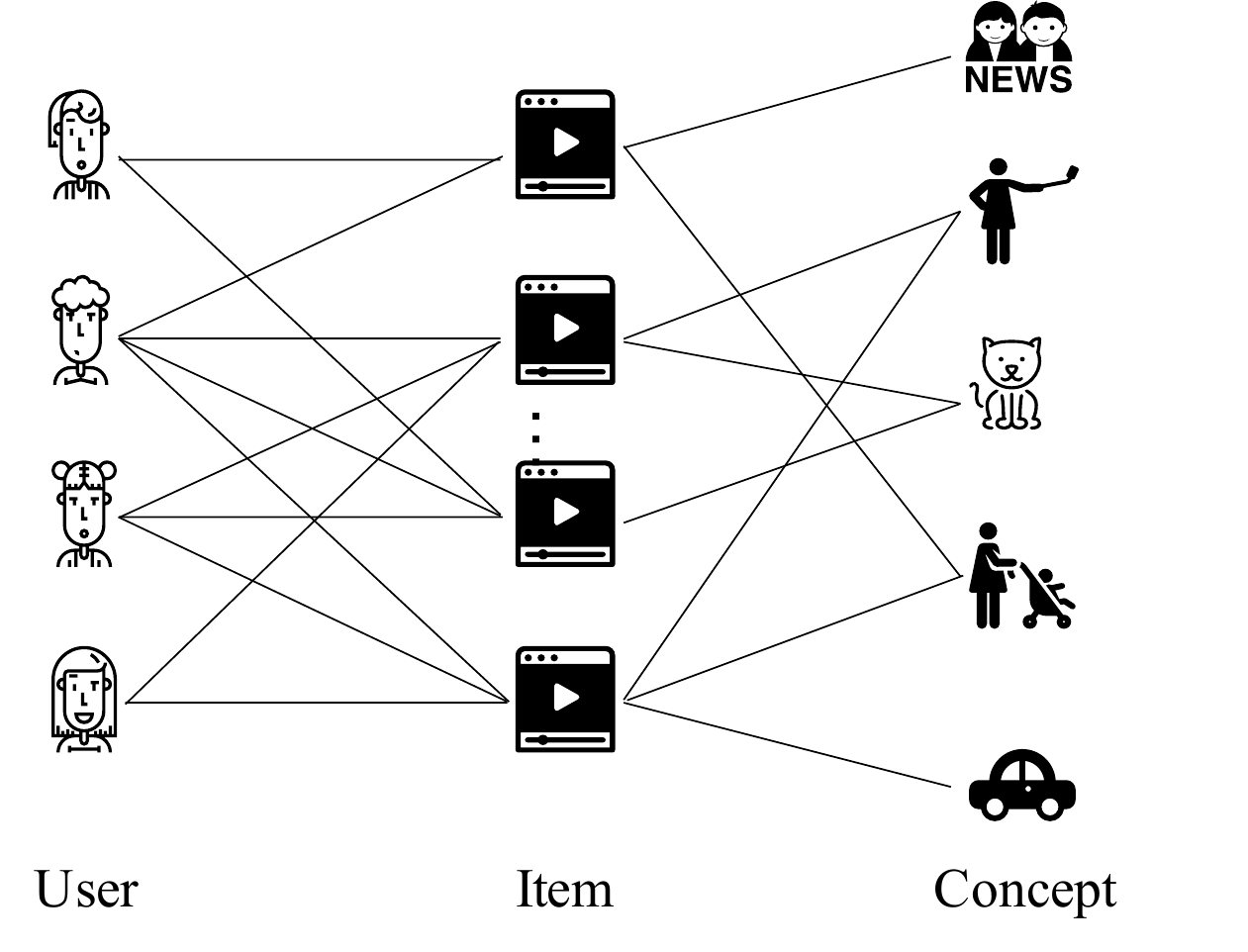}
\centering
\caption{The tripartite heterogeneous graph which is composed of three disjoint vertex  sets: users, items, and concepts, such that no two graph vertices within the same set are adjacent.}
\label{fig:graph}
\end{figure}

In this section, we present a concept-aware denoising graph neural network for micro-video recommendation. Specifically, the proposed \baby aims to derive the representations for users and micro-videos by propagating the conceptual semantics to both relevant users and micro-videos. Illustrated in Figure~\ref{fig:graph}, \baby consists of three phases: \textit{warm-up propagation, graph denoising} and \textit{preference refinement}. In the following, we describe each phase following their usage order in \baby.

\subsection{Warm-Up Propagation}\label{ssec:warmup}
Firstly, after concept extraction for micro-videos, we form a tripartite heterogeneous graph $\mathcal{G}=(\mathcal{V},\mathcal{E})$, where $\mathcal{V}$ and $\mathcal{E}$ are the node set and the edge set respectively. A node $o\in\mathcal{V}$ could be a user $u$, a micro-video $m$ or a concept $c$. Moreover, there are two kinds of edges in $\mathcal{E}$, which connect a user with a micro-video when the user clicks the latter, and a micro-video with a concept when the concept is extracted from the micro-video. Figure~\ref{fig:graph} illustrates an example of this tripartite heterogeneous graph. Since we plan to inject the conceptual information into the user and item representations, at the beginning, we perform a series of graph convolution operations in terms of concepts in $\mathcal{V}$. 

\begin{figure*}[htbp]
\centering
\includegraphics[width=\textwidth]{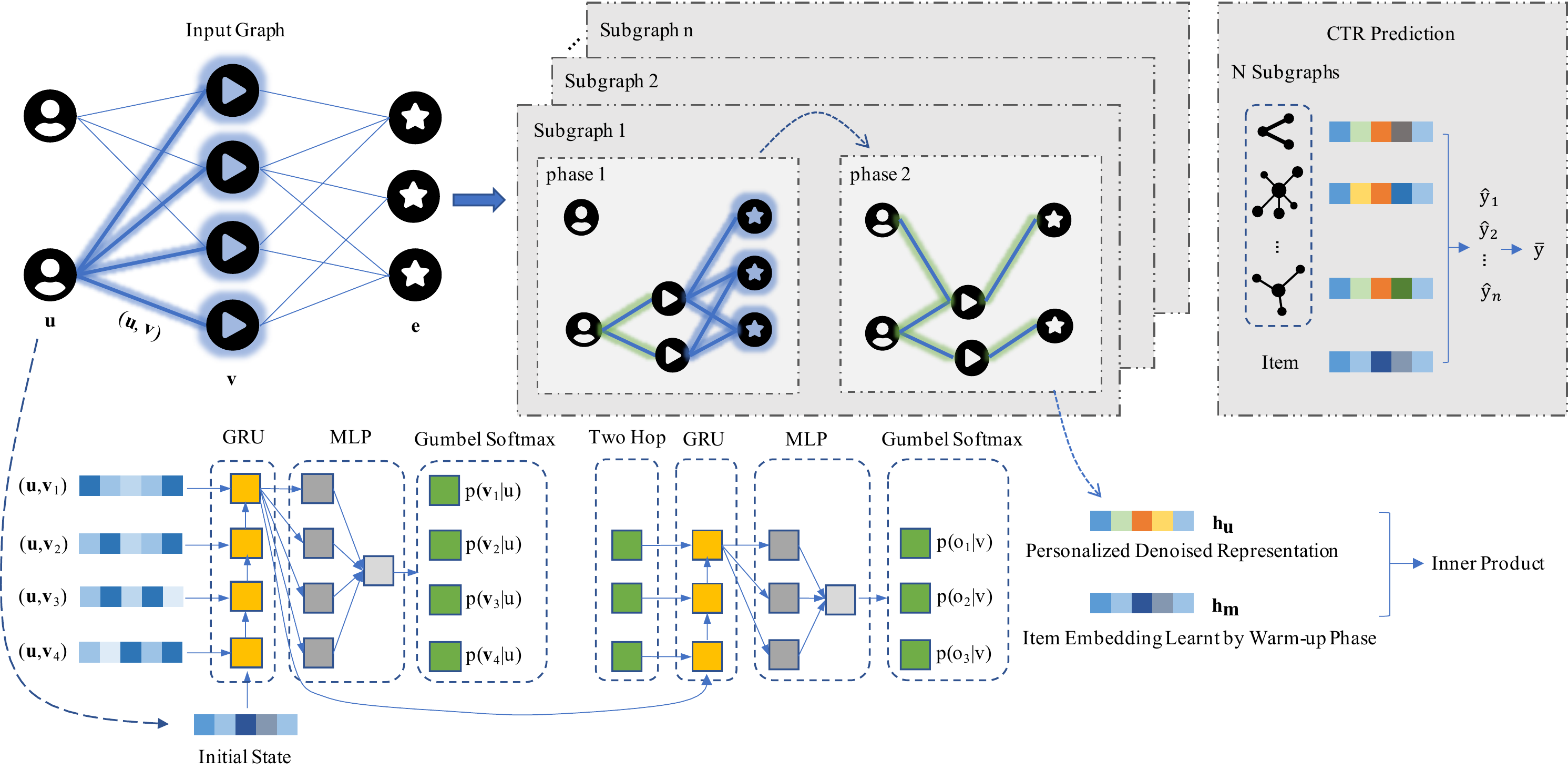}
\centering
\caption{The network architecture of our proposed CONDE.}
\label{fig:CONDE}
\end{figure*}

Specifically, given a micro-video $m$ with a set of concept neighbors $\mathcal{C}_m$, the hidden feature vector $\mathbf{h}_m$ of micro-video $m$ can be derived by aggregating the embeddings of its concept neighbors: $\mathbf{h}_m=AGG(\mathbf{e}_m,\{\mathbf{e}_c,\forall c\in\mathcal{C}_m\})$, where $\mathbf{e}_m,\mathbf{e}_c$ are the embedding vector for micro-video $m$ and its concept neighbor $c$ respectively. Given the simplicity and effectiveness, we choose graph attention network\cite{GAT/iclr2018} to instantiate $AGG(\cdot)$ function as follows:
\begin{align}
\mathbf{h}_m&=\sigma(\sum_{c\in\mathcal{C}_m}\alpha_c\mathbf{W}\mathbf{e}_c)\label{eqn:warm-up-video}\\
\alpha_c&=\frac{\exp(\sigma(\mathbf{a}^{\top}[\mathbf{e}_m\|\mathbf{e}_c]))}{\sum_{i\in\mathcal{C}_m}\exp(\sigma(\mathbf{a}^{\top}[\mathbf{e}_m\|\mathbf{e}_i]))}\label{eqn:attn}
\end{align}
where $\alpha_c$ is the attention weight indicating the importance of concept $c$, $\sigma$ is the LeakyReLU activation and $\|$ is the vector concatenation operation. It is expected that $\mathbf{h}_m$ could encode semantic information regarding the content of micro-video $m$. 

Following similar operation, we then derive the hidden preference vector $\mathbf{h}_u$ of user $u$ by using micro-video neighbors $\mathcal{M}_u$ of the latter: $\mathbf{h}_u=AGG(\mathbf{e}_u,\{\mathbf{h}_j,\forall j\in\mathcal{M}_u\})$, where $\mathbf{e}_u$ is the embedding vector of user $u$, and $\mathbf{h}_j$ is the hidden feature vector of micro-video neighbor $j$ derived in Equation~\ref{eqn:warm-up-video}. It is clear now that we propagate the semantic information first from concepts to micro-videos, and then from micro-videos to users. 

So far, the whole propagation process is purely concept driven. Note that not all micro-videos are well expressed by their concept neighbors. Plus, the collaborative signals are not exploited during this process. Hence, we then apply another convolution operation for micro-videos by aggregating their user neighbors back. Specifically, the hidden feature vector $\mathbf{h}_m$ of micro-video $m$ is updated with its user neighbor set $\mathcal{U}_m$: $\mathbf{h}_m=AGG(\mathbf{h}_m,\{\mathbf{h}_u,\forall u\in\mathcal{U}_m\})$. In this sense, we can propagate the conceptual semantics across the user-video bipartite part further, enriching the feature representation of micro-videos, especially the long-tail ones.

\subsection{Graph Denoising}\label{ssec:denoising}

It is intuitive that different users would be interested in different aspects of the micro-video content. Plus, there are inevitably some noisy concepts and false clicks to distract the effective representation learning for users. Although we have an attention mechanism in Equation~\ref{eqn:attn}, the irrelevant and noisy information still complicates the learning process. Here, we introduce a user-oriented denoising process to filter the micro-videos and the concepts in a breadth-first search fashion. 

Specifically, we adopt a gated recurrent unit (GRU) as the compositer to firstly identify the relevance between user $u$ and each of her micro-video neighbors as follows: 
\begin{align}
\mathbf{f}_{u,m}=GRU(\mathbf{h}_u,\mathbf{h}_m), \forall m\in\mathcal{M}_u\label{eqn:gru-u}
\end{align}
where $\mathbf{h}_u$ works as the initial state of GRU, and $\mathbf{f}_{u,m}$ encodes the relevant information. Then, we perform the neighbor denoising as a sampling without replacement to retain only $n$ micro-video neighbors: $\mathcal{M}^{\star}_u=Den(\{\mathbf{f}_{u,m},\forall m\in\mathcal{M}_u\})$ and $|\mathcal{M}^{\star}_u|=n$. These $n$ micro-videos are expected to convey the user's preference more precisely. In detail, we can utilize a fully-connected layer with a softmax function to derive the likelihood of retaining each micro-video neighbor $m$ as follows:
\begin{align}
s_{u,m}=\frac{\exp(\mathbf{w}^{\top}\mathbf{f}_{u,m})}{\sum_{i\in\mathbf{M}_u}\exp(\mathbf{w}^{\top}\mathbf{f}_{u,i})}
\end{align}
where $\mathbf{w}$ is the parameter vector for graph denoising. Note that instead of attention mechanism, the denoising process produces discrete selections, which are not
differentiable for model learning. Hence, we choose Gumbel-Softmax\cite{gumbel} to instantiate $Den(\cdot)$ for differentiable discrete sample generation: 
\begin{align}
\pi_{u,m}=\frac{\exp((\log(s_{u,m})+\epsilon_m)/\tau)}{\sum_{i\in\mathbf{M}_u}\exp((\log(s_{u,m})+\epsilon_m)/\tau)}\label{eqn:gumbel}
\end{align}
where $\epsilon_{\ast}=-\log(-\log(x))$ and $x$ is i.i.d sampled from Uniform($0,1$). The temperature parameter $\tau$ controls sharpness of the likelihood distribution. When $\tau$ is small, Equation~\ref{eqn:gumbel} produces a multi-modal distribution. On the contrary, when $\tau$ is large, the resultant distribution is nearly equivalent to a uniform one. 

Afterwards, we continue the above denoising process for the user's two-hop neighbors via each micro-video $m$ in $\mathcal{M}^{\star}_u$. Note that we only consider the concept of neighbors of $m$: $\mathcal{N}_{u,m}=\mathcal{C}_m$. We have tried mixing user and concept neighbors in this second phase, which turns out to be inferior to using concept nodes only. We speculate that putting different types of nodes in GRU simultaneously could complicate the model learning, since heterogeneous information is difficult to be fused together. 

Then, we still utilize the same compositer in the first-hop denoising to calculate the relevance information:
\begin{align}
\mathbf{f}_{u,m,v}=GRU(\mathbf{f}_{u,m},\mathbf{h}_v), \forall v\in\mathcal{N}_{u,m}\label{eqn:gru-m}
\end{align}
where $\mathbf{h}_v$ is the representation of two-hop neighbor $v$. Here, the concept embedding $\mathbf{e}_c$ is used (\ie $\mathbf{h}_v=\mathbf{e}_c$). Also, when we consider the neighboring users together, when $v$ is user $u'$, the hidden preference vector $\mathbf{h}_{u'}$ is used instead (\ie $\mathbf{h}_v=\mathbf{h}_{u'}$). Then, we apply the similar denoising process for $\mathcal{N}_{u,m}$ and obtain $n$ two-hop neighbors for user $u$ via micro-video $m$: $\mathcal{N}^{\star}_{u,m}=Den(\{\mathbf{f}_{u,m,i},\forall i\in\mathcal{N}_{u,m}\})$ and $|\mathcal{N}^{\star}_{u,m}|=n$.

We can see that the whole denoising process is performed sequentially by starting with each user. Since the GRU-based compositor is initialized with the user preference vector of the warm-up phase, the user's neighborhood in $\mathcal{G}$ is reshaped to express the user's preference more precisely in a subgraph $\mathcal{G}^{\star}_u$, where the nodes are $\{u,\mathcal{M}^{\star}_u,\mathcal{N}^{\star}_{u,m},\forall m\in\mathcal{M}^{\star}_u\}$, and the corresponding edges between them in $\mathcal{G}$ are remained\footnote{Note that we can continue the denoising process for three-hop neighbors and more. The computation cost is very high hence we leave it for further work.}.

\subsection{Preference Refinement}\label{ssec:preference}
Now, it is straightforward to refine the hidden preference representation for each user based on the corresponding $\mathcal{G}^{\star}_u$. Specifically, for each micro-video $m$ in $\mathcal{M}^{\star}_u$, we firstly refine the hidden feature vector $\mathbf{h}^{\star}_m$ for $m$: $\mathbf{h}^{\star}_m=AGG(\mathbf{e}_m,\{\mathbf{h}_v,\forall v\in\mathcal{N}^{\star}_{u,m}\})$. Here, as to $\mathbf{h}_v$, we adopt the same setting as in Equation~\ref{eqn:gru-m}. Then, we refine the hidden preference vector $\mathbf{h}^{\star}_u$: $\mathbf{h}^{\star}_u=AGG(\mathbf{e}_u,\{\mathbf{h}^{\star}_m,\forall m\in\mathcal{M}^{\star}_{u}\})$. At last, the ranking score for a micro-video $m$ w.r.t. user $u$ is calculated as follows:
\begin{align}
\hat{y}_{u,m}=sigmoid(\mathbf{h}^{\star\top}_u\mathbf{h}_m)
\label{eqn:prediction}
\end{align}
where $\mathbf{h}_m$ is the hidden feature vector derived in the warm-up phase for micro-video $m$.

\subsection{Model Optimization}\label{ssec:optimization}
Besides the embedding vectors and parameters for graph convolution, the key of \baby is to perform graph denoising in a personalized way to enhance the user preference learning. Following the work in~\cite{robust/ICML2019}, for each user $u$, we can generate $k$ subgraphs $\{\mathcal{G}^{\star}_{u,1},\dots,\mathcal{G}^{\star}_{u,k}\}$ by repeating the denoising process $k$ times. After that, we calculate the cross-entropy loss with each subgraph $\mathcal{G}^{\star}_{u,i}$ as follows:
\begin{align}
\mathcal{L}_{u,m,i}=y_{u,m}\log(\hat{y}_{u,m})+(1-y_{u,m})\log(1-\hat{y}_{u,m})
\end{align}
where $y_{u,m}$ is the ground truth of micro-video $m$ clicked by user $u$, $\hat{y}_{u,m}$ is the corresponding ranking score calculated with subgraph $\mathcal{G}^{\star}_{u,i}$. Finally, the total loss function is as follows:
\begin{align}
\mathcal{L}=\sum_u\sum_m\sum_i\mathcal{L}_{u,m,i}+\lambda\left\lVert \Theta \right\rVert^{2}
\end{align}
where $\left\lVert \Theta \right\rVert^{2}_2$ represents $L2$ regularization over model parameters and $\lambda$ is the corresponding coefficient. 

\paratitle{Discussion.} After model training, we follow the order of \textit{warm-up}~$\to$~\textit{graph denoising}~$\to$~\textit{preference refinement} for each user and generate the corresponding subgraph $\mathcal{G}^{\star}_{u}$ and user preference vector $\mathbf{h}^{\star}_u$. We then maintain these preference vectors in memory for online service. There is no need to re-run the graph denoising and preference refinement each time, since we believe that the learnt model is effective in capturing the user's preference. The marginal change in the subgraph structure would introduce ignorable variations. Actually, all micro-video hidden feature vectors could be reused in the warm-up propagation phase for each user, by calculating $\mathbf{h}_m$ beforehand.

\section{Experiments}
We evaluate our proposed \baby on two real-world datasets, hoping to answer the following three research questions: \textbf{RQ1} How does \baby perform compared with state-of-the-art graph-based methods? \textbf{RQ2} How do different components affect \baby's performance? \textbf{RQ3} Can \baby provide reasonable explanations about user preferences towards items?

In what follows, we first present the experimental settings in detail, (\ie the datasets, baselines, evaluation protocols, and the parameter settings), then compare our model's performance with state-of-the-art methods, followed by answering the above three questions.

\subsection{Dataset Description}\label{ssec:dataset}
\paratitle{Micro-Video} dataset is collected from a popular large scale Chinese micro-video sharing platform. It serves to record and share videos taken by a user and matches content to targeted consumers with the help of a recommender system. The recommendation model serves in a streaming setting, where users can swipe up on the touch screen to watch the next video one by one.

To be as consistent as possible with the real-world online training setting, we construct a dataset by randomly sampling $100K$ users and their watched micro-videos over a period of two days. After filtering the users and micro-videos with no click behavior, the dataset consists of $82,193$ users and $784,863$ clicked micro-videos, which is referred to as \textit{Micro-Video} dataset in the rest of this paper. The user behaviors in first-day are used for training. We then shuffle and split the second-day's data equally for validation and testing respectively. 

\paratitle{Amazon Electronics}~\cite{amazondata} is a widely used benchmark whose task is to predict whether a user will write a review for a target item given her historical reviews (in English). Initially we wanted to find a public micro-video recommendation dataset in English but failed. So we have to turn to data from other fields. Surprisingly, \baby works out quite well on other types of tasks and thus proving its robustness. Here, we select the first 70\% of historical interactions as the training set. The last 10\% interactions are used as the testing set. The remaining 20\% is used as the validation set for hyper-parameter selection. We refer to this benchmark as the \textit{Amazon} dataset.

\begin{table}
    \caption{Statistics of the datasets.}
    \label{tab:dataStats}
    \begin{tabular}{ccccc}
      \toprule
      \textbf{Dataset}&\textbf{\#User}&\textbf{\#Item}&\textbf{\#Concept}&\textbf{\#Record}\\
      \midrule
      Micro-Video &82,193 &784,863 &415,796 &9,455,804 \\
      Amazon Electronics  & 192,403 &63,001 &185,894 & 1,612,381 \\
    \bottomrule
  \end{tabular}
\end{table}

\subsection{Data Preprocessing}
As for the Micro-Video dataset, we build a concept inventory by merging several proprietary knowledge bases and internal video tagging systems. Specifically, the resultant concept inventory contains about $2.32$M named entities and semantic key phrases, covering a wide spectrum of topics in Micro-Video. The average length of a concept is $4$. The concepts are extracted from textual information of a micro-video including comment, caption and hashtag by performing a longest match. As the user generated content contains a lot of Internet buzzwords and slangs, we filter the corpus first by removing stop words, punctuation and emoji. After concept extraction, we calculate the TF-IDF score for each concept and micro-video pair, and retain the informative ones with a threshold filtering. In total, we obtained $415,796$ unique concepts. For the Amazon dataset, we utilize the TextBlob toolkit\footnote{\url{https://github.com/sloria/TextBlob}} to extract the noun phrases as the concepts from the reviews written for each item. Similarly, a TF-IDF weighting scheme is adopted to filter noisy concepts. The detailed statistics of the two datasets are reported in Table~\ref{tab:dataStats}.

\begin{table*}[htbp]
  \centering
  \caption{Performance comparison between methods on the Micro-Video and Amazon dataset. The * indicates the best performance of the baselines. The best results are highlighted in boldface.}
  \label{tab:performance_comparison}
    \begin{tabular}{ccccccccc}
    \toprule
    \multirow{2}{*}{Method}&
    \multicolumn{4}{c}{Micro-Video}&\multicolumn{4}{c}{Amazon}\cr
    \cmidrule(lr){2-5} \cmidrule(lr){6-9}
    &AUC&NDCG@5&HIT@5&MAP@5&AUC&NDCG@5&HIT@5&MAP@5 \cr
    \midrule
    PinSage & 0.7254 & 0.8922 & 0.2772 & 0.2610     & 0.6957 & 0.7688 & 0.7185  & 0.6109\cr
    GraphSage & 0.7139 & 0.8847 & 0.2750 & 0.2580   & 0.6961 & 0.7703 & 0.7204 & 0.6125\cr
    NIA-GCN & 0.7203 & 0.8657 & 0.2717 & 0.2517      &0.6903
& 0.7647 & 0.7174 & 0.6049 \cr
    \midrule
    GAT   & 0.7345* & 0.9102* & 0.2811* & 0.2668*      & 0.7009* & 0.7734* & 0.7210* & 0.6162* \cr
    HAN   & 0.7251 & 0.8988 & 0.2782 & 0.2630      & 0.6984 & 0.7724 & 0.7204 & 0.6144\cr
    \midrule
    \baby &\textbf{0.7952}&\textbf{0.9140}&\textbf{0.2890}&\textbf{0.2747}&\textbf{0.7064}&\textbf{0.7792 } & \textbf{0.7266} &\textbf{0.6226}\cr
    \bottomrule
    \end{tabular}
    
\end{table*}

\subsection{Baselines.}
\begin{itemize}
    \item \textbf{GraphSage}~\cite{graphSage/nips17} is a state-of-the-art method of GCN that aggregates neighborhood information into center nodes. In particular, it considers the structure information as well as the distribution of node features in the neighborhood. For a fair comparison, we form two homogeneous graphs based on user-item interaction. 
    \item \textbf{PinSage}~\cite{pinSage/KDD18} is designed to employ GraphSAGE on user-item graph, which performs efficient, localized convolutions by sampling the neighborhood around a node and dynamically constructing a computation graph. In this work, we employ two graph convolution layers as suggested in the user-item interaction graph.
    \item \textbf{GAT}~\cite{GAT/iclr2018} introduces a multi-head attention between each node and its neighbors to calculate aggregate coefficients. 
    \item \textbf{HAN}~\cite{HAN/www2019} is a heterogeneous graph neural network with attention mechanism based on GAT, which deals with heterogeneous nodes via different meta paths to aggregate neighborhood information. We modify the model for top-N recommendation. In detail, we make use of four types of meta-paths: user-item-user, user-item-concept-item-user, item-concept-item, and item-user-item. Therefore we construct four homogeneous graphs from the original graph for model learning.
    \item \textbf{NIA-GCN}~\cite{nia-gcn} models the interactions between neighbors with element-wise products by a bi-linear neighborhood aggregator. The Euclidean distance between users and items with their positive neighbors are added in loss function. The experimental setting is the same as the original NIA-GCN paper. This is the uptodate state-of-the-art solution in the paradigm of GNN.
\end{itemize}
For those network embedding models (\ie GraphSage, GAT and HAN), we also utilize Equation~\ref{eqn:prediction} to calculate the ranking score for a fair comparison.

\subsection{Experimental Settings}

With the extracted concepts, a heterogeneous concept-aware graph is built by connecting the items to the associated concepts and users to their consumed items. 
Recall that we aim to propagate semantic concept information via graph convolution to represent users and micro-videos for preference learning. 
In the warm-up propagation phase, we derive coarse-level representations for micro-videos and users following the aggregation order of concept $\to$ micro-video $\to$ user. Note that the concept-aware graph is not perfect but contains lots of noisy information. . At last, we again perform the same convolution as in the warm-up propagation phase over the subgraph to refine the user representation for recommendation. 

\eat{
In our experiments, we try different denosing strategies, such as take both user and concept into consideration in the second stage, which turns out to be inferior to our proposed method. There is a speculation that putting different types of nodes in GRU simultaneously could produce more noise, leading to unstable prediction. After denoising, the nodes are prepared for information aggregating. For the aggregation strategy, we also test and verify different plans. The result shows that taking both user and concept nodes into consideration in 2-hop aggregation works out best. 
}

\paratitle{Evaluation Metrics.} For performance evaluation metrics, we choose AUC (Area Under Curve), HIT@5, NDCG@5, MAP@5 which are widely adopted in many related works~\cite{sigir21:qin,sigir21:tian}. Note that we use UAUC here, which means AUC score for each user, and we calculate the mean of all users' AUC scores as the final result. Meanwhile, We treat all unobserved interactions as the negative instances when reporting performance. It is worthwhile to mention that the positive and negative instances are highly imbalanced. As for model training, the ratio of true and negative samples in real production scenarios is approximate to $1$:$1$. Hence, we also choose the same number of unobserved items for each user in evaluation.

\paratitle{Hyperparameter Settings.} Following PinSage~\cite{pinSage/KDD18}, we perform mini-batch training and edge sampling for better training efficiency. According to the average number of concepts that an item contains, we randomly sample a subgraph for training by randomly picking $p$ neighbors in every epoch. We set $p = 40$ for the Micro-Video dataset and $p = 100$ for the Amazon dataset after grid search. The proposed \baby will select a fixed number of neighbors out of this subgraph (\ie $n<p$). For fair comparison, after grid search, we fix the embedding size and all hidden sizes to be $128$ and $50$ for the Micro-Video and the Amazon dataset respectively across different methods in comparison. As for \baby, we firstly apply Word2Vec~\cite{word2vec} to pre-train the word embeddings over the textual information of items in the Micro-Video dataset. Here, the textual information associated with a micro-video is merged as a single document. For the Amazon dataset, we use pre-trained GloVe word embeddings\footnote{https://github.com/stanfordnlp/GloVe}. The concept embeddings are fixed as the averaged embedding of their constituent words, and are not further fine-tuned during the model training. Recall that the temperature parameter $\tau$ in Equation~\ref{eqn:gumbel} could affect the sample probabilities significantly. The higher the temperature, the smoother the sampling probability is. Starting from a higher temperature, the model can gather information from more neighbors in the first few minibatches. In our experiments, the temperature starts at a high value and gradually anneal to a small one as follows:
$$\tau = \tau_0 \times exp(-\eta x)$$
where $\tau_0$ is the initial temperature to start with, $\eta$ is the hyperparameter, and $x$ is the number of minibatches. We set $\tau_0$ to be $10$, and $\eta$ to be $0.0002$ and $0.001$ for the Micro-Video and the Amazon dataset respectively.

\subsection{RQ1: Performance Comparison}
Table~\ref{tab:performance_comparison} reports the overall performance compared with baselines. For each model, we run five times and report the average results. From the results, we make the following observations.
\begin{itemize}
    \item \baby consistently yields the best performance on all datasets. This result demonstrates the superiority of our concept-aware denoising framework for recommendation.

    \item It is no surprise that methods based on a bipartite graph like PinSage~\cite{pinSage/KDD18} and NIA-GCN~\cite{nia-gcn} consistently yields poor performance on all evaluations, which indicates that learning user and item representations simply on a bipartite graph is not adequate.

    \item The results of GraphSage~\cite{graphSage/nips17} show that homogeneous graphs with directly connected items and users can not make up for the negative impact of noisy information. Attention mechanism contributes a lot to GAT~\cite{GAT/iclr2018} performance improvement and HAN~\cite{HAN/www2019} are more powerful in information utilization and aggregation  with item-concept connection when constructing item-item, user-user graph homogeneous graph. 

    \item However, according to our experiments, none of them ever achieves the best result, which verifies the advantages of the denoising optimization process of our proposed model.
\end{itemize}

\paratitle{Long-Tail Performance.}
As to Micro-Video dataset, we divide the entire item space into hot and long-tail items by click frequency. We set 50 as a threshold and find that long-tail items take up a percentage of $96.0\%$ in the entire item space. However, the majority of user-item interactions in the test set are dominated by hot items, leaving only $34.48\%$ records for long-tail ones to share. We split the test set into two parts according to click frequency$=50$ and do the experiments on them separately. From Table~\ref{tab:long-tail} we can observe that the performance of each model for hot items is always much higher than that of long-tail items, which indicates the distributions of hot and long-tail items are inconsistent. What's more, the improvements of \baby over the baselines are remarkable. Therefore we can draw a conclusion that \baby shows its strong robustness on both hot and long-tail items, and proves its ability to learn better features in the entire space for different datasets and applications.

\begin{table}[h]
  \centering
  \caption{Performance comparison between the methods with \baby and other baselines on Micro-Video dataset. ``Hot'' represents hot items in the test set, ``Long-tail'' represents long-tail items in the test set.}
  \label{tab:long-tail}\resizebox{\linewidth}{!}{
    \begin{tabular}{ccccccccc}
    \toprule
    \multirow{2}{*}{Method}&
    \multicolumn{2}{c}{AUC}&\multicolumn{2}{c}{NDCG@2}&\multicolumn{2}{c}{HIT@2}&\multicolumn{2}{c}{MAP@2}\cr
    \cmidrule(lr){2-3} \cmidrule(lr){4-5} \cmidrule(lr){6-7} \cmidrule(lr){8-9}
    &Hot&Long-tail&Hot&Long-tail&Hot&Long-tail&Hot&Long-tail\cr
    \midrule
    PinSage &0.7524 &0.6231&0.9176&0.7135&0.2470&0.3332&0.2358&0.2929 \cr
    GraphSage &0.7662&0.6436&0.9234&0.7298&0.2473&0.3378&0.2369&0.2985 \cr
    GAT & 0.7710 & 0.6526  & 0.9409  & 0.7134  &  0.2505  & 0.3406 & 0.2413 &  0.3018\cr
    HAN &0.7585&0.6419&0.9320&0.7107&0.2487&0.3366&0.2387&0.2970 \cr
    NIA-GCN &0.7606 &0.6354&0.9118&0.6694&0.2443&0.3307&0.2332&0.2883\cr
    
    \midrule 
    \baby &\textbf{0.8331}&\textbf{0.6924}&\textbf{0.9445}&\textbf{0.7587}&\textbf{0.2552}&\textbf{0.3471}&\textbf{0.2552}&\textbf{0.3122}\cr
    
    \bottomrule
    \end{tabular}}
\end{table}

\subsection{RQ2: Model Analysis}

\paratitle{Ablation Study.} We conduct ablation studies on the Micro-Video dataset for each proposed component to justify its effectiveness. For each setting, we run 5 times and report the average results. 

Note that our graph is tripartite, which is composed of three disjoint vertex sets: users, items, and concepts, such that no two graph vertices within the same set are adjacent. Therefore we perform the denoising algorithm in a two-phase strategy. Starting from a user, GRU first picks some candidate micro-videos out of its neighbors in the first phase. Then for every remaining micro-video, the same GRU picks some candidate concepts out of its neighbors in the second phase. The whole process can be split into two phases, hence we can do some ablation studies as follows: 

For better comparison, we split the denoising process into two phases and remove either phase with a random module, which randomly selects some micro-videos instead of denoising, to check if denoising operation really makes some difference. 

\begin{small}
\begin{table}[H]
  \centering
  \caption{Ablation study on the Micro-Video dataset.}
  \label{tab:abla}\resizebox{\linewidth}{!}{
    \begin{tabular}{ccccc}
    \toprule
    \multirow{2}{*}{Method}&
    \multicolumn{4}{c}{Micro-Video}\cr
    \cmidrule(lr){2-5} 
    &AUC&NDCG@5&HIT@5&MAP@5\cr
    \midrule
    random phase 1 & 0.7870 & 0.9106 & 0.2877 & 0.2729 \cr
    random phase 2 & 0.7854  & 0.9090  & 0.2866 & 0.2714 \cr
    random phase 1 \& 2 & 0.7780   & 0.9087  & 0.2856 & 0.2704 \cr
    denoising phase 1  & 0.7780 & 0.9042 & 0.2849 & 0.2689 \cr
    denoising phase 2  & 0.7874 & 0.9106 & 0.2871 & 0.2718 \cr
    denoising phase 1 \& 2 &\textbf{0.7952}&\textbf{0.9140}&\textbf{0.2890}&\textbf{0.2747} \cr
    \bottomrule
    \end{tabular}}
\end{table}
\end{small}
From the results shown in Table~\ref{tab:abla}, we make the following observations: 1) Cut out either denoising phase is detrimental to the final result. 2) The second phase of denoising plays a more important role. 3) The best performance appears when two phases work together, indicating that two phases should work together to enhance representation learning. This is reasonable since the user's representation is refined in terms of the relevant micro-videos whose representations are also dependent on the relevant concepts. This pipeline convolution nature for user representation learning requires us to denoising the irrelevant information in each step. 

\paratitle{Number of Remaining Neighbors.} Given the graph to denoise, we choose a fixed number of nodes for information aggregation. To concentrate on crucial points, we mainly discuss the concept denoising in this paper. Note that we have also tested a different number of one-hop neighbors for each user, where we decide to pick 6 micro-videos for every user out of her 10 neighbors. Here, we test the number of remaining concepts amongst $\{6,10,20,30\}$ out of $40$. The number of subgraphs is fixed to be four in this case.

From Table~\ref{tab:number of neighbors} we can draw a conclusion that if we leave too many neighbors, noisy information will still exist and harm the model training. Yet if we leave too few neighbors, the model is likely to be over-denoised and miss much important information. In general, this parameter is not very sensitive for model performance. We could suggest using a relatively small value to enhance computation efficiency. 

\begin{table}[h]
    \centering
    \setlength{\intextsep}{0mm}
    \caption{Performance over different $k$ values.}
    \label{tab:number of neighbors}\resizebox{\linewidth}{!}{
    \begin{tabular}{ccccc}
    \toprule
    \multirow{2}{*}{Remaining Neighbors}&
    \multicolumn{4}{c}{Micro-Video}\cr
    \cmidrule(lr){2-5} 
    &AUC&NDCG@5&HIT@5&MAP@5\cr
    \midrule
    n=6 & 0.7942 & 0.9082 & 0.2883 & 0.2733  \cr
    n=10 &\textbf{0.7952}&\textbf{0.9140}&\textbf{0.2890}&\textbf{0.2747} \cr
    n=20 & 0.7913 & 0.9112 & 0.2875 & 0.2723 \cr
    n=30 & 0.7924 & 0.9131 & 0.2885 & 0.2739  \cr
    \bottomrule
    \end{tabular}}
  \end{table}

\paratitle{Number of Subgraphs}
Apparently, the number of subgraphs is also an important parameter to \baby. Similarly, we test the number of subgraphs amongst $\{2,4,6\}$, and Table~\ref{tab:subgraph} indicates that adding the number of subgraphs improves the performance further. However the improvement of evaluation metrics can not keep the same pace with the resultant increasing training time. As the training cost is proportional to the number of subgraphs and the model can not obtain further remarkable performance gain by adding more subgraphs, we use $k=4$ in this paper.

\begin{table}[H]
    \centering
    \caption{Performance over different $G$ values.}
    \label{tab:subgraph}\resizebox{\linewidth}{!}{
    \begin{tabular}{ccccc}
    \toprule
    \multirow{2}{*}{Number of Subgraphs}&
    \multicolumn{4}{c}{Micro-Video}\cr
    \cmidrule(lr){2-5} 
    &AUC&NDCG@5&HIT@5&MAP@5\cr
    \midrule
    G=2  & 0.7898 & 0.9071 & 0.2866 & 0.2710 \cr
    G=4  & \textbf{0.7952} & 0.9140 & \textbf{0.2890} & \textbf{0.2747}  \cr
    G=6 & 0.7941 & \textbf{0.9143} & 0.2887 & 0.2741 \cr
    
    \bottomrule
    \end{tabular}}
  \end{table}

\paratitle{Impact of Different Temperatures.} We also test \baby's sensitivity to different temperatures. Specifically speaking, we set different starting temperatures and annealing ratios, and the model performs best when $\tau=10$ and $\eta=2e-4$ on Micro-Video dataset as shown in Table~\ref{tab:tau}.

\begin{table}[H]
    \centering
    \caption{Performance over different temperature settings. }
    \label{tab:tau}\resizebox{\linewidth}{!}{
    
    \begin{tabular}{ccccc}
    \toprule
    \multirow{2}{*}{Different Starting Temperature}&
    \multicolumn{4}{c}{Micro-Video}\cr
    \cmidrule(lr){2-5} 
    &AUC&NDCG@5&HIT@5&MAP@5\cr
    \midrule
    10 ($\eta=2e-4$)  & \textbf{0.7952} & \textbf{0.9140} & \textbf{0.2890} & \textbf{0.2747}  \cr
    100 ($\eta=5e-4$) & 0.7875  & 0.9103 & 0.2870 &  0.2718  \cr 
    1000 ($\eta=1e-3$) & 0.7859  & 0.9098 & 0.2870  & 0.2717  \cr 
    \bottomrule
    \end{tabular}}
  \end{table}

\begin{figure*}[h]
    \centering
  \begin{subfigure}[b]{0.4\textwidth}
    \includegraphics[width=0.95\textwidth]{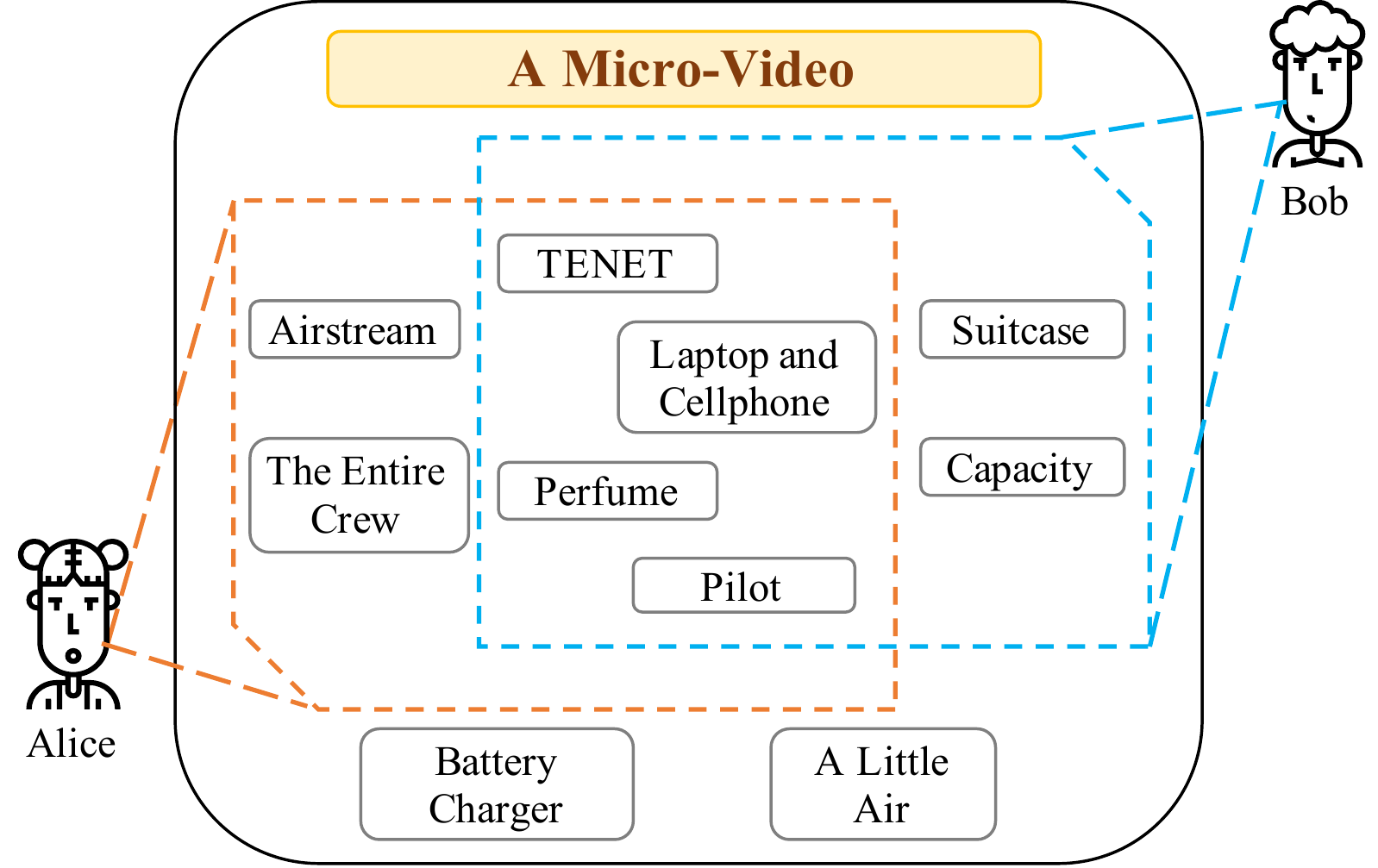}
    \label{explaina1A}
    
  \end{subfigure}
   \begin{subfigure}[b]{0.4\textwidth}
    \includegraphics[width=0.9\textwidth]{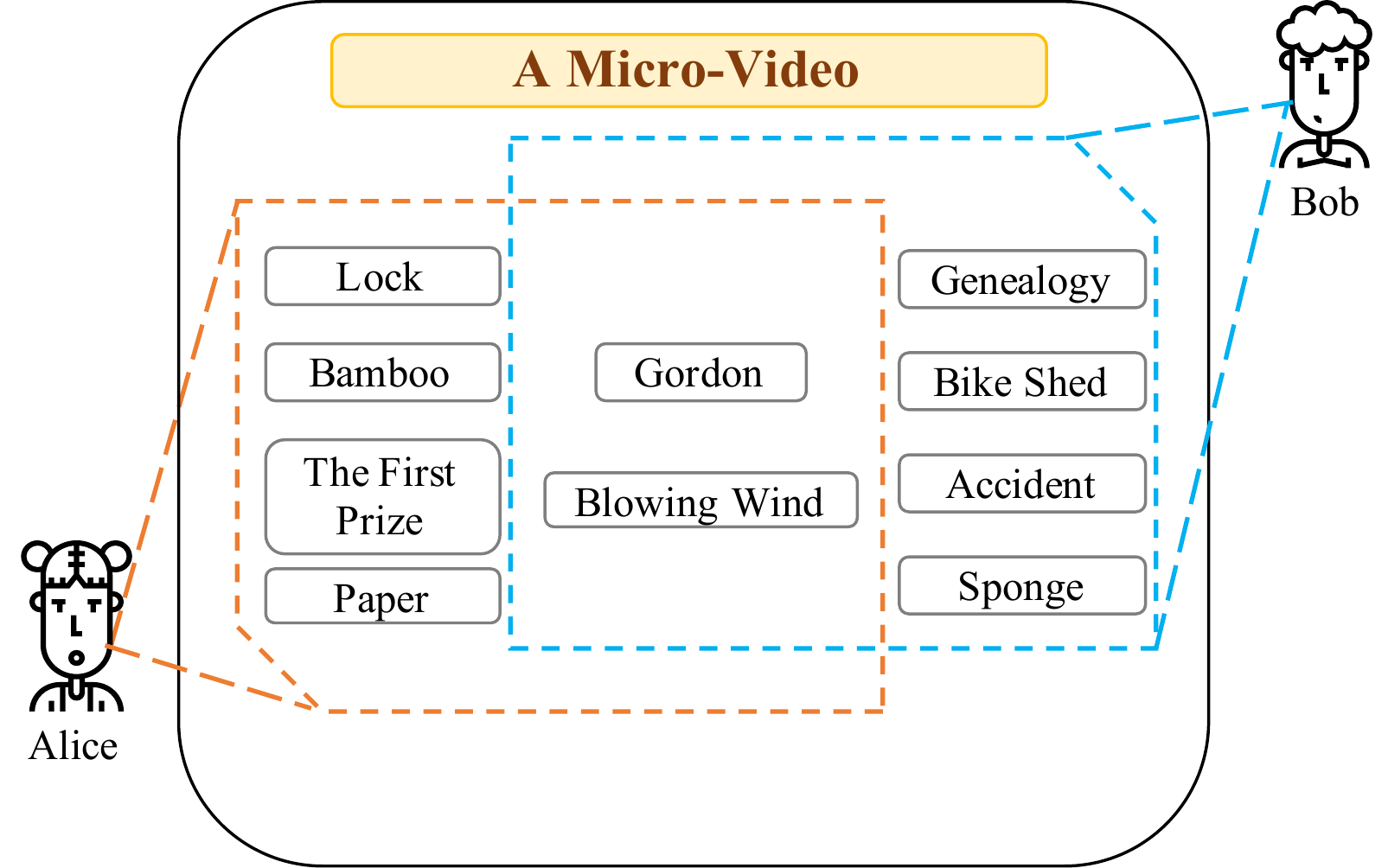}
   
    \end{subfigure}
     \caption{Examples of two users focusing on different points of the same item.}
     \label{explain1B}
    \hspace{10mm}
    
    \begin{subfigure}[b]{0.8\textwidth}
    \includegraphics[width=0.9\textwidth]{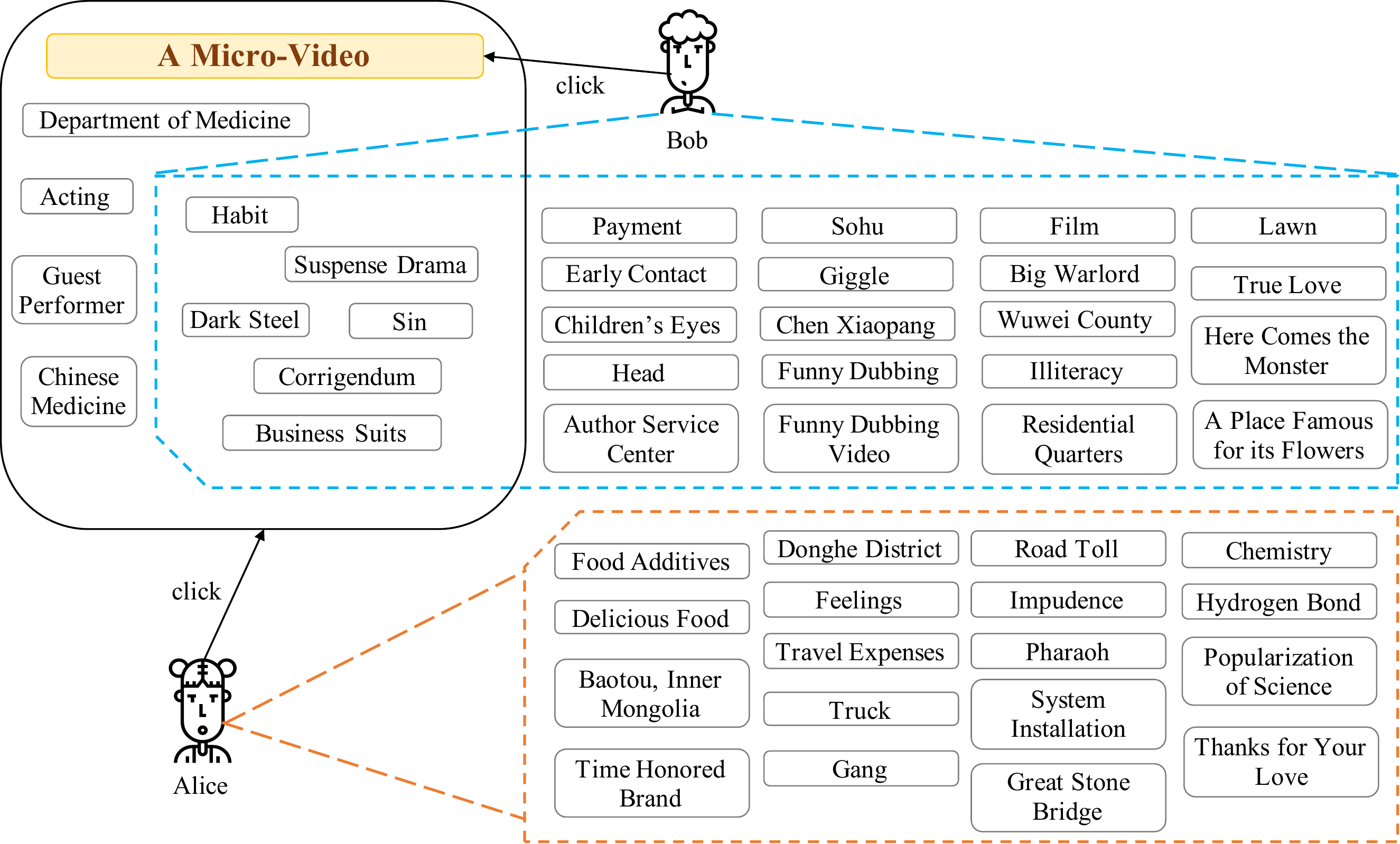}
    \label{explain2A}
    \centering
    \end{subfigure}
    \caption{Two users interacted with the same item.}
\label{fig:Explainability}
\end{figure*}

\subsection{RQ3: Explainability Analysis}
We further investigate whether \baby can filter out noise and retain more meaningful information. To better visualize the results, we pick some users and their corresponding tripartite subgraphs, with items and concepts connected to each other. As ID is not explicit to describe an item, we use each micro-video's concept neighbors instead, to facilitate better understanding of its content.

In order to demonstrate the explainability of \baby, we first choose two user pairs, each of which share at least one same relevant item after performing denoising, according to \baby's result. As demonstrated in Figure~\ref{explain1B}, the concepts of the item are split out by orange and blue lines are the relevant ones after 2-phase denoising for the two users respectively. Here we use the Micro-Video dataset and translate every concept in English for easy understanding. We can observe that although two users have the same relevant item, they focus on different points of interest. Figure~\ref{explain1B} further indicates that a micro-video itself can include various points of interest.

What's more, we find another pair of users who interacted with the same item, and this item is deleted for one user but kept for the other by \baby. We compare the related item and all remaining concepts of these two users in Figure~\ref{fig:Explainability}. The result shows that there is an obvious divergence between their preferences. Overall, these results suggest that \baby is able to discover informative yet relevant concepts out of noisy ones, besides enhancing the recommendation performance, but also support recommendation explainability.

\section{Conclusion }
In this paper, we propose a novel \textbf{con}cept-aware \textbf{de}noising graph neural network (named \baby) to address information redundancy challenges. To make full use of the textual information in micro-video recommendation scenarios, we extract concepts from captions and the comments associated with micro-videos. In this way, we can form a tripartite heterogeneous graph by connecting user, micro-video, and concept nodes. As a user will interact with hundreds of micro-videos and a micro-videos will receive lots of comments everyday, neither all micro-videos nor all concepts of a micro-video can reflect users' preference precisely. Hence, we propose a personalized denoising methodology to derive user and item representations via graph neural network. 
The extensive experiments over a large micro-video dataset and a traditional E-Commerce dataset in two different languages demonstrate the effectiveness of \baby. By abundant ablation studies with different experiment settings, the proposed \baby has proven its excellent potential to reflect the user's preference.

\begin{acks}
    This work was funded by Kuaishou and National Natural Science Foundation of China (No.~61872278). Chenliang Li is the corresponding author.
\end{acks}

\bibliographystyle{ACM-Reference-Format}
\balance
\bibliography{refer}

\end{document}